\documentclass[10pt,twocolumn,twoside]{opticajnl}
\journal{opticajournal} 
\setboolean{shortarticle}{true}

\usepackage{xcolor}

\usepackage{dblfloatfix}
\usepackage{mathrsfs}
\usepackage{placeins}
\usepackage{braket}

\title{High-Fidelity Prediction of Perturbed Optical Fields using Fourier Feature Networks}
\author[1]{Joshua R. Jandrell}
\author[1,*]{Mitchell A. Cox}
\affil[1]{School of Electrical and Information Engineering, The University of the Witwatersrand, Johannesburg 2050, South Africa}
\affil[*]{Mitchell.Cox@wits.ac.za}

\begin{abstract}
Predicting the effects of physical perturbations on optical channels is critical for advanced photonic devices, but existing modelling techniques are often computationally intensive or require exhaustive characterisation. We present a novel data-efficient machine learning framework that learns the perturbation-dependent transmission matrix of a multimode fibre. To overcome the challenge of modelling the resulting highly oscillatory functions, we encode the perturbation into a Fourier Feature basis, enabling a compact multi-layer perceptron to learn the mapping with high fidelity. On experimental data from a compressed fibre, our model predicts the output field with a 0.995 complex correlation to the ground truth, improving accuracy by an order of magnitude over standard networks while using 85\% fewer parameters. This approach provides a general tool for modelling complex optical systems from sparse measurements.
\end{abstract}

\setboolean{displaycopyright}{false}

\begin{document}
\maketitle
Novel transmission matrix engineering devices, such as \textit{all-optical wavefront modulators} \cite{Resisi2020_wavefront_shaping, Shekel2024_og_piano_tut} harness controlled physical perturbations, like bends in multi-mode optical fibre, to transform the propagating (input) field, $U$, into a new (output) field $V$. These transformations have applications in modulation; efficient high-fidelity error correction; and optical computing \cite{Shekel2024_og_piano_tut, Qiu2024_temporal_shaper}. However, their development is limited by our ability to accurately characterise and simulate them. Current work relies on pre-recorded data or "online" real-time measurements for configuration and experimentation \cite{Shekel2024_og_piano_tut, Qiu2024_temporal_shaper}.

Although many theoretical models (like the modified Helmholtz and non-linear Schroedinger equations) can describe the effects of a perturbation on the propagating field; they require computationally intensive numerical solutions and precise knowledge and/or control of the fibre's physical properties \cite{Shekel2024_og_piano_tut}. Furthermore, these models do not accurately match physical observations due to microscopic irregularities in the geometry and refractive index of the fibre \cite{Shekel2024_og_piano_tut, Ye2025_seefibre}.

A similar challenge exists when attempting to perform robust error correction for optical communications \cite{Ye2025_seefibre} and high-fidelity imagining like single fibre endoscopy \cite{wen_endoscope_2023}. If the transformation $U\rightarrow V$ is known or predicted, then it is easier to reverse. But, again, we are limited by our ability to accurately model and characterize systems without extensive measurements.

Deep learning networks, such as multi-layer perceptrons (MLPs), present a natural solution to this problem as they can capture more nuance than purely theoretical models \cite{Ye2025_seefibre} and can thus interpolate between data points, greatly reducing number of experimental measurements required \cite{goel_referenceless_2023}. These are commonly called physics informed neural networks (PINNs).

However, when learning the mapping $f(U, d) \rightarrow V$ where $d$ is some known perturbation, a standard MLP faces a fundamental dilemma when modelling a complex-valued optical field, $z \in \mathbb{C}$. The choice of representation creates a trade-off. In its polar form, $z = |z|e^{i\phi_z}$, the phase $\phi_z$ exhibits jump discontinuities at intervals of $2\pi$, which prevents effective learning via gradient descent. In its Cartesian form, $z = a_z + ib_z$, the real and imaginary components are continuous and differentiable. But this latter approach confronts a different and more subtle obstacle: \textit{spectral bias}, the documented tendency of MLPs to struggle with learning high-frequency functions \cite{tancik_fourier_2020}. In many physical systems, such as optical modes propagating through a perturbed fibre, the phase evolves rapidly with respect to the perturbation. This causes the real and imaginary components, $a_z = |z|\cos(\phi_z)$ and $b_z = |z|\sin(\phi_z)$, to become highly oscillatory functions that the network cannot accurately approximate. Consequently, the network's spectral bias prevents it from accurately approximating these components, leading to significant errors in the predicted field.

\medskip

Here, we present a method to overcome this limitation. We equip an MLP with \textit{Fourier features}: a set of sinusoidal basis functions proposed by Tancik \textit{et al.} \cite{tancik_fourier_2020} to counteract spectral bias in general machine learning applications like image processing. Instead of only feeding the network the normalized perturbation $\tilde{d}$, we provide a higher-dimensional feature vector, $\gamma(\tilde{d})$, as an additional input:
\begin{equation}
    \begin{split}
        \gamma(\tilde{d}) = \big[&\cos(\omega_1 \tilde{d}),~\sin(\omega_1 \tilde{d}),~\dots, \\
        &\cos(\omega_M \tilde{d}), ~\sin(\omega_M \tilde{d})\big]
    \end{split}
    \label{eq:fourierFeatures}
\end{equation}
where $\{\omega_m\}_{m=1}^M$ is a collection of fixed frequencies. This encoding simplifies the MLP's task from learning a periodic function to learning the appropriate linear combination of the provided basis sinusoids in $\gamma(\tilde{d})$. 

We follow a philosophy of physics-informed problem decomposition: the model should not learn what can be computed directly. Rather than predicting the output field $V$, our networks predict the perturbation-dependent transmission matrix, $T(d)$. An arbitrary complex field, $C$, can be expressed as a linear superposition of orthogonal modes $\Psi_n$, $C \approx \sum_{n=1}^{N} c_n \Psi_n(s)$, which can be represented by a vector of modal coefficients $\ket{c} = [c_1, \dots, c_N]^T$ \cite{pinnell2020modal}. The input-output relationship is then a simple matrix multiplication,
\begin{equation}
    \ket{v}=T(d)\ket{u}.
\end{equation}
Since this is a linear operation, we only learn the mapping $f(d) \rightarrow T(d)$. For a constant input mode $\ket{u_j}$, the task reduces to learning a single column of the TM: $f(d) \rightarrow\ket{T_j(d)}=\ket{v}$. This significantly reduces problem complexity. The complex vector $\ket{v} \in \mathbb{C}^N$ is represented as a real-valued vector $\ket{y} \in \mathbb{R}^{2N}$ by concatenating its real and imaginary parts:
\begin{equation}
    \ket{y}=\left[\text{Re}\{\ket{v}\}^T, \text{Im}\{\ket{v}\}^T\right]^T.
    \label{eq:realVector}
\end{equation}

\begin{figure}[t]
    \centering
    \includegraphics[width=0.99\linewidth]{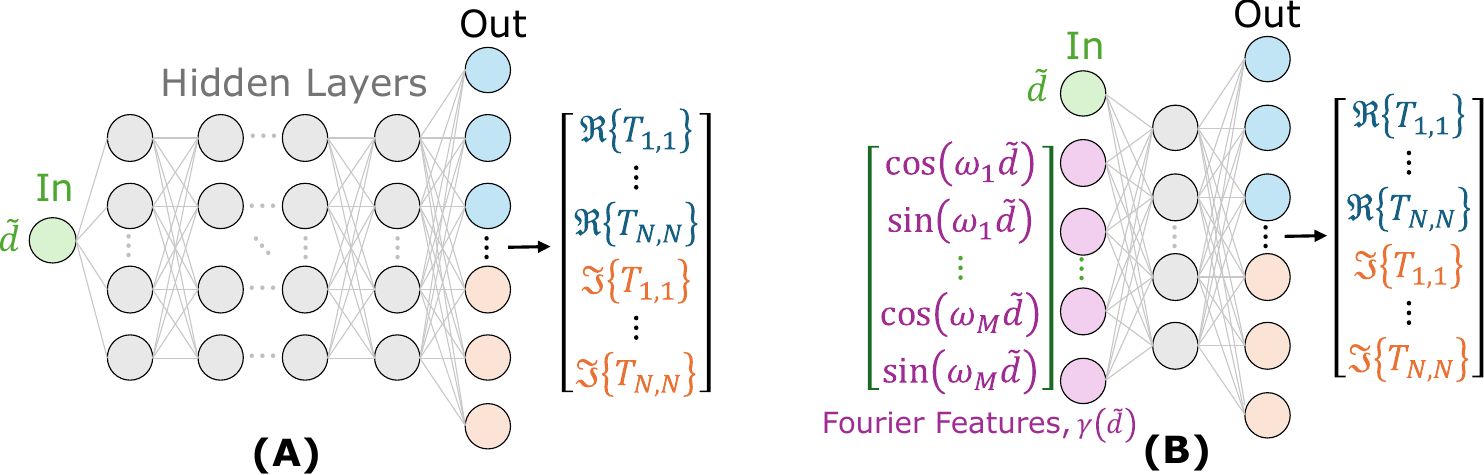}
    \caption{Neural network architectures used predict the perturbation dependent transmission matrix $T(\tilde{d})=\Re\{T\}+i\Im\{T\}$. \textbf{(A)} The standard MLP requires a deep network of hidden layers while \textbf{(B)} FNET takes additional inputs $\gamma(\tilde{d})$ but has fewer parameters.}
    \label{fig:nets}
\end{figure}

We tested this approach on the dataset from Matthès et al.~\cite{matthes_learning_2021}, which records the complex TM of a multi-mode fibre under a compressive deformation, $d \in [0, 70]\,\text{\textmu m}$. We fixed the input field to $U = \Psi_1 =\text{OAM}_{0,1}$ (a Gaussian) and learned the $N=20$ perturbation-dependent output modal weights for the basis of ten clockwise and ten anti-clockwise polarized OAM modes, resulting in a 40-neuron output size. For training data, we used 18 evenly spaced sample matrices from the dataset, with the remaining 18 used as unseen test cases to assess each model's ability to interpolate.

\medskip

The core task, therefore, was to train a network to predict the perturbation-dependent column of the transmission matrix. To rigorously evaluate our proposed Fourier feature approach, we compared three architectures for this task, each representing a distinct modelling philosophy (depicted conceptually in Fig.~\ref{fig:nets}).

First, a baseline Multi-Layer Perceptron (MLP) serves to demonstrate the fundamental challenge. Due to the spectral bias discussed previously, we expect this standard architecture to fail at fitting the highly oscillatory target function. Our baseline used a deep $16 \times 20$-neuron (360 total neurons) architecture and a $\text{hardtanh}(x)$ activation function.

Second, we benchmarked against a Sinusoidal Representation Network (SIREN) \cite{NEURIPS2020_sirens}. A SIREN is a specialised architecture also designed to overcome spectral bias, but it does so \textit{implicitly}. By using a periodic activation function, $\sin(\omega_0 x)$, throughout the network, it attempts to discover the necessary sinusoidal components as part of the training process. This provides a critical point of comparison: in contrast to our FNET, which is explicitly given a fixed frequency basis, the SIREN must learn an appropriate basis from the data. Our SIREN used three 20-neuron hidden layers and $\omega_0=6\pi$, matching the approximate centre frequency of the physical system. The SIREN has 100 neurons in total.

Third, our proposed Fourier Feature Network (FNET) combines a standard MLP backbone with the explicit, high-frequency basis provided by the Fourier feature vector, $\gamma(\tilde d)$. This architecture is intentionally shallow, using only a single 20-neuron hidden layer with a $\text{hardtanh}(x)$ activation. The FNET had 60 neurons in total.

The FNET's Fourier feature vector was constructed using a linear progression of frequencies, $\{\omega_m=2\pi m\}_{m=1}^5$, a choice informed by the approximately $\pm3$ phase wraps observed in the data. The Nyquist-Shannon sampling theorem provides a useful heuristic for selecting these frequencies: for a normalized deformation $\tilde{d} \in [0, 1]$ sampled at $L$ points, any frequency $\omega_m$ exceeding the limit $\omega_\text{max} = \pi L$ risks being misinterpreted as an aliased, lower-frequency component, which would degrade interpolation performance.

To ensure a fair comparison, all models were trained using an Adam optimiser (learning rate 0.001) for 200 epochs to minimise the mean squared error (MSE) loss, $\mathscr{L}_\text{MSE}$. The SIREN was given the specialised initialisation required for effective training, with weights selected from the uniform distribution $\mathcal{U}[-\sqrt{6/n_\text{in}}/\omega_0, \sqrt{6/n_\text{in}}/\omega_0]$, where $n_\text{in}$ is the layer's input dimensionality \cite{NEURIPS2020_sirens}.

The stability of the FNET's training is a direct consequence of its simple architecture, which reframes the learning problem. Instead of approximating a highly oscillatory function from scratch -- a task with a notoriously non-convex loss landscape prone to undesirable local minima -- the network's primary task is to find the optimal linear combination of a pre-defined sinusoidal basis. This simplifies the optimisation to a quasi-linear regression problem with a much smoother loss landscape, ensuring reliable convergence. This stability is a prerequisite for more advanced strategies, such as so-called curriculum learning. 

While $\mathscr{L}_\text{MSE}$ on the Cartesian components is effective for establishing the correct phase, it provides no explicit constraint on the predicted magnitude. This is a known trade-off: small, uncorrelated errors in the real and imaginary parts can combine to produce larger, non-monotonic errors in the amplitude. To address this anticipated effect, we introduce a loss term specifically for the amplitude, leading to a combined loss function:
\begin{equation}
   \mathscr{L}_\text{Combined}=\frac{\alpha\mathscr{L}_\text{MSE}(\ket{\hat y}, \ket{y})+\beta\mathscr{L}_\text{MSLE}(\ket{\hat v}, \ket{v})}{\alpha+\beta},
   \label{eq:combinedLoss}
\end{equation}
where $\mathscr{L}_\text{MSLE}$ is the Mean Squared Logarithmic Error, given by:
\begin{equation}
    \mathscr{L}_\text{MSLE}(\ket{\hat v}, \ket{v})=\frac{1}{N}\sum_{n=1}^N(\log(1+|\hat v_n|)-\log(1+|v_n|))^2
    \label{eq:msleLoss}
\end{equation}
and $\alpha:\beta=1:10$ are weighting factors. We adopt a curriculum: the model is first trained for 100 epochs with $\mathscr{L}_\text{MSE}$ to establish a good phase-space solution, followed by 100 epochs with $\mathscr{L}_\text{Combined}$ to refine the amplitude prediction.

\begin{table*}[tb]
\centering
\caption{\bf Model Performance on Unseen Test Data (Averaged over 50 runs)}
\begin{tabular}{cc|cc|ccc}
\hline
Model & Training Regime & $\ket{\hat{v}}$ (MAE) & $\ket{\hat{v}}$ ($\rho$) & $\hat{V}$ (MAE) & $\hat{V}$ ($\rho$) & SSIM \\
\hline\hline
MLP & MSE only & $8.44\text{e}{-2}\angle0.40\pi$ & $0.953\angle0.369\pi$ & $3.25\text{e}{-3}\angle0.3843\pi$ & $0.939\angle0.369\pi$ & $0.9186$\\
& curriculum & $3.62\text{e}{-2}\angle0.44\pi$ & $0.928\angle0.411\pi$ & $1.56\text{e}{-3}\angle0.4206\pi$ &$0.921\angle0.410\pi$ & $0.9605$\\
\hline
SIREN & MSE only & $4.07\text{e}{-2}\angle0.26\pi$ & $0.926\angle0.130\pi$ & $1.33\text{e}{-3}\angle0.2466\pi$ & $0.891\angle0.138\pi$ & $0.9698$\\
& curriculum & $4.33\text{e}{-2}\angle0.29\pi$ & $0.920\angle0.122\pi$ & $1.41\text{e}{-3}\angle0.2545\pi$ & $0.873\angle 0.131\pi$ & $0.9694$\\
\hline
FNET & MSE only & $9.34\text{e}{-3}\angle\mathbf{0.10}\boldsymbol{\pi}$ & $\mathbf{0.998}\angle0.063\pi$ & $3.53\text{e}{-4}\angle{\mathbf{0.0835}\boldsymbol{\pi}}$ & $\mathbf{0.996}\angle{0.063\pi}$ & $0.9968$\\
& curriculum & $\mathbf{7.29\text{e}{-3}}\angle0.12\pi$ & $0.997\angle\mathbf{0.060}\boldsymbol{\pi}$ & $\mathbf{3.21\text{e}{-4}}\angle0.0855\pi$ & $0.995\angle\mathbf{0.060}\boldsymbol{\pi}$ & $\mathbf{0.9974}$ \\
\end{tabular}
  \label{tab:train-1-results}
\end{table*}

\begin{figure}[tb]
    \centering
    \includegraphics[width=1.0\linewidth]{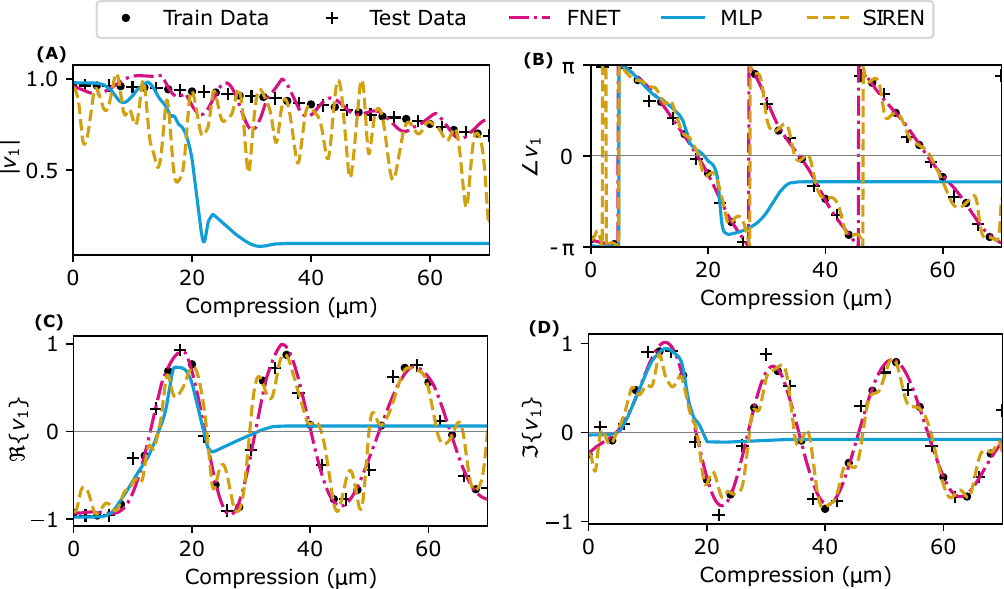}
    \caption{Predictions of the \textbf{(A)} magnitude, \textbf{(B)} phase, \textbf{(C)} real, and \textbf{(D)} imaginary components for the first mode weight $\ket{v_1}$ with respect to compressive perturbation $d$.}
    \label{fig:mse-tracks}
\end{figure}

\begin{figure}[tb]
    \centering
    \includegraphics[width=1.0\linewidth]{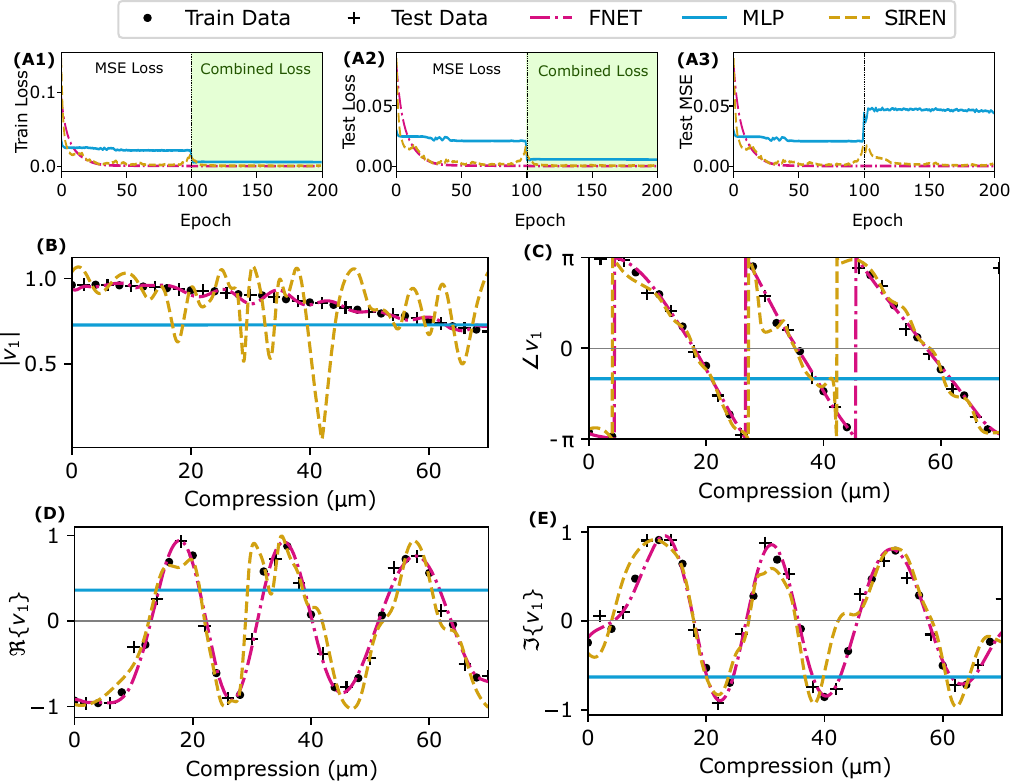}
    \caption{\textbf{(A1)} Training and \textbf{(A2)} testing loss for curriculum learning with \textbf{(A3)} MSE test loss as a reference. The FNET now predicts \textbf{(B)} the magnitude of the first mode weight $\ket{v_1}$ without introducing ripples while still fitting \textbf{(C)} phase, \textbf{(D)} real, and \textbf{(E) }imaginary components.}
    \label{fig:curriculum-tracks}
\end{figure}

We now present the results, beginning with a direct comparison of the three architectures under standard training. Figure~\ref{fig:mse-tracks} shows the comparative performance for predicting a single modal weight, $\ket{v_1}$, using the standard MSE loss. 

The baseline MLP (\raisebox{3\height}{\includegraphics[width=1em]{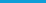}}) partially fits the data for $d\in[0,20]$ \textmu m but fails to capture the oscillatory behavior. The SIREN (\raisebox{3\height}{\includegraphics[width=1em]{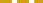}}) captures periodicity but introduces significant high-frequency noise, particularly in the amplitude. This is a result of implicit internal frequencies above the sampling limit. In contrast, the FNET (\raisebox{3\height}{\includegraphics[width=1em]{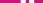}}) converges smoothly and accurately models all components without high-frequency noise. The FNET is more consistent and robust because its architecture decouples the periodic basis (fixed features, $\gamma(\tilde{d})$) from the linear combination (learned weights) to create a simpler optimization problem.

Figure~\ref{fig:curriculum-tracks} illustrates the results of the curriculum training. The FNET architecture successfully accommodates the change in loss function, reducing amplitude fluctuations without compromising the underlying phase prediction. The MLP struggles to satisfy both criteria, while the SIREN's reliance on purely sinusoidal activations limits its ability to make the smooth, aperiodic adjustments needed for amplitude control. In contrast, the FNET's architecture, which combines linear transformations with periodic basis functions, provides the necessary flexibility.

The performance of each model on unseen test data is quantified in Tab.~\ref{tab:train-1-results}, averaged over 50 runs. The FNET significantly outperforms the other models across all metrics. Notably, its high correlation $|\rho|$ remains consistent when moving from modal weights ($\ket{\hat{v}}$) to the full field ($\hat{V}$), indicating uniform accuracy across all modes. 

\begin{figure}[t]
    \centering
    \includegraphics[width=1\linewidth]{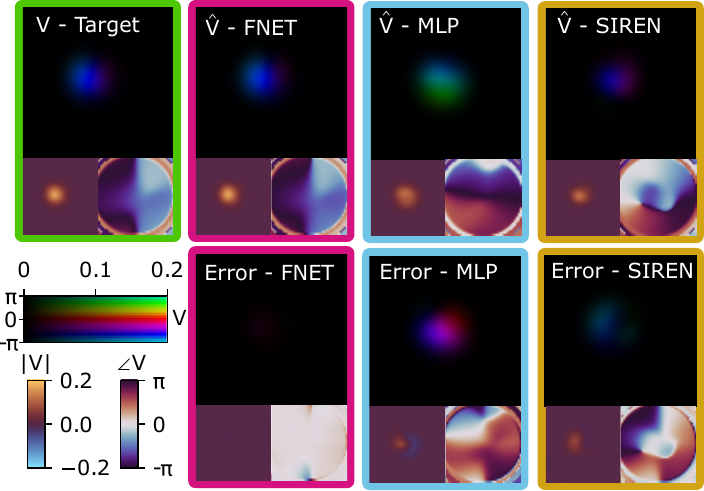}
    \caption{Predicted field $\hat{V}$ and error (difference to target) for an unseen compression $d=42\,\text{\textmu m}$.}
    \label{fig:feilds}
\end{figure}

The curriculum regime improved the FNET's amplitude MAE by 22\% with only a minor trade-off in phase accuracy. The final, high-fidelity reconstruction of an unseen output field is shown in Fig.~\ref{fig:feilds}. The FNET prediction is visually indistinguishable from the ground truth, whereas the baseline MLP fails to capture the correct spatial structure or phase. The FNET predicted these fields with a mean complex Pearson correlation of $\bar\rho=0.995\angle0.06\pi$ (absolute phase), while the MLP achieved only $\bar\rho=0.911\angle0.41\pi$, despite the FNET using 85\% fewer parameters.

\medskip

The FNET's capacity to optimize for multiple criteria makes it suitable for physics-informed neural networks, where physics-based loss functions can be incorporated. Because the curriculum was implemented with minimal hyperparameter tuning, there is likely an opportunity for further optimization. There is also ample scope to apply FNET to more complicated frequency profiles or systems with multiple interdependent perturbations. Future work can also investigate the construction of the feature vector, $\gamma$. For example, in cases where a perturbation follows a known non-linear mapping $g(\tilde{d})$, the transformed feature vector $\gamma\circ g(\tilde{d})$ may prove effective. Another consideration is that the FNET has direct access to the value of $\tilde{d}$, which is ambiguous after the first SIREN layer due to the many-to-one mapping of its activation function. This could be addressed by re-introducing $\tilde{d}$ after each SIREN layer, a topic for future work.

\medskip

In conclusion, we have demonstrated that encoding a perturbation parameter as a set of Fourier features enables a standard MLP to model highly oscillatory complex optical fields with remarkable accuracy. By factorizing the physics problem—learning the perturbation-dependent transmission matrix rather than the full field-to-field map—and using a curriculum learning strategy, our FNET model substantially outperforms baseline MLP and SIREN architectures. This approach alleviates the need for lookup tables or empirical measurements, while being several orders of magnitude more computationally efficient than complex simulations. It is general and robust, suggesting its applicability to a wide range of problems in optics and other domains involving oscillatory systems.

\begin{backmatter}
\bmsection{Funding}
National Research Foundation.
\bmsection{Acknowledgment} Thank you Bob and Geraud N. Tasse for the fascinating discussions. 
\bmsection{Disclosures} The authors declare no conflicts of interest.

\bmsection{Data availability} Models were trained on the dataset generated by Matthès \textit{et. al.} 
\cite{matthes_learning_2021}. Pre-trained model and python code for constructing and training FNETS is available on \hyperlink{https://github.com/WitsOCLab/Fourier_Feature_Networks_for_High-Fidelity_Prediction_of_Perturbed_Optical_Fields}{GitHub}\footnote{\url{https://github.com/WitsOCLab/FNETs}}.
\end{backmatter}
\bibliography{sample}

\begin{thebibliography}{10}
\newcommand{\enquote}[1]{``#1''}

\bibitem{Resisi2020_wavefront_shaping}
S.~Resisi, Y.~Viernik, S.~M. Popoff, and Y.~Bromberg, \enquote{Wavefront shaping in multimode fibers by transmission matrix engineering,} {\protect\JournalTitle{APL Photonics}} \textbf{5} (2020).

\bibitem{Shekel2024_og_piano_tut}
R.~Shekel, K.~Sulimany, S.~Resisi, \emph{et~al.}, \enquote{Tutorial: How to build and control an all-fiber wavefront modulator using mechanical perturbations,} {\protect\JournalTitle{JPhys Photonics}} \textbf{6} (2024). The OG fiber piano tutorial paper.

\bibitem{Qiu2024_temporal_shaper}
T.~Qiu, H.~Cao, K.~Liu, \emph{et~al.}, \enquote{Spectral-temporal-spatial customization via modulating multimodal nonlinear pulse propagation,} {\protect\JournalTitle{Nature Communications}} \textbf{15} (2024).

\bibitem{Ye2025_seefibre}
Z.~Ye, T.~Zhao, and W.~Xia, \enquote{Seeing through multimode fibers using real-valued intensity transmission matrix with deep learning,} {\protect\JournalTitle{Optics Express}} \textbf{33}, 16222 (2025).

\bibitem{wen_endoscope_2023}
Z.~Wen, Z.~Dong, Q.~Deng, \emph{et~al.}, \enquote{Single multimode fibre for in vivo light-field-encoded endoscopic imaging,} {\protect\JournalTitle{Nature Photonics}} \textbf{17}, 679--687 (2023). Publisher: Nature Publishing Group.

\bibitem{goel_referenceless_2023}
S.~Goel, C.~Conti, S.~Leedumrongwatthanakun, and M.~Malik, \enquote{Referenceless characterization of complex media using physics-informed neural networks,} {\protect\JournalTitle{Optics Express}} \textbf{31}, 32824 (2023). Publisher: Optica Publishing Group.

\bibitem{tancik_fourier_2020}
M.~Tancik, P.~Srinivasan, B.~Mildenhall, \emph{et~al.}, \enquote{Fourier {Features} {Let} {Networks} {Learn} {High} {Frequency} {Functions} in {Low} {Dimensional} {Domains},} in \emph{Advances in {Neural} {Information} {Processing} {Systems},}  vol.~33 (Curran Associates, Inc., 2020), pp. 7537--7547.

\bibitem{pinnell2020modal}
J.~Pinnell, I.~Nape, B.~Sephton, \emph{et~al.}, \enquote{Modal analysis of structured light with spatial light modulators: a practical tutorial,} {\protect\JournalTitle{Journal of the Optical Society of America A}} \textbf{37}, C146--C160 (2020).

\bibitem{matthes_learning_2021}
M.~W. Matthès, Y.~Bromberg, J.~De~Rosny, and S.~M. Popoff, \enquote{Learning and {Avoiding} {Disorder} in {Multimode} {Fibers},} {\protect\JournalTitle{Physical Review X}} \textbf{11} (2021). ArXiv: 2010.14813 Publisher: American Physical Society.

\bibitem{NEURIPS2020_sirens}
V.~Sitzmann, J.~Martel, A.~Bergman, \emph{et~al.}, \enquote{Implicit neural representations with periodic activation functions,} in \emph{Advances in Neural Information Processing Systems,}  vol.~33 H.~Larochelle, M.~Ranzato, R.~Hadsell, \emph{et~al.}, eds. (Curran Associates, Inc., 2020), pp. 7462--7473.

\end{thebibliography}
\bibliographyfullrefs{sample}
\end{document}